\documentclass{emulateapj}
\def\s2n{S^{\prime}/N}

\shorttitle{}
\shortauthors{Padoan, Haugb\o lle, and Nordlund}

\begin{document}
\title{A Simple Law of Star Formation}

\author{Paolo Padoan}
\affil{ICREA \& ICC, University of Barcelona, Marti i Franqu\`{e}s 1, E-08028 Barcelona, Spain;
ppadoan@icc.ub.edu}
\author{Troels Haugb\o lle}
\affil{Centre for Star and Planet Formation, University of Copenhagen, \OE stervoldgade 5-7., DK-1350,
Copenhagen, Denmark; haugboel@nbi.dk}
\author{\AA ke Nordlund}
\affil{Centre for Star and Planet Formation and Niels Bohr Institute, University of Copenhagen,
Juliane Maries Vej 30, DK-2100, Copenhagen,
Denmark; aake@nbi.dk}

\begin{abstract}

We show that supersonic MHD turbulence yields a star formation rate (SFR) as low as observed in molecular clouds (MCs), for characteristic values of the free-fall time divided by the dynamical time, $t_{\rm ff}/t_{\rm dyn}$, the alfv\'{e}nic Mach number, ${\cal M}_{\rm a}$, and the sonic Mach number, ${\cal M}_{\rm s}$. Using a very large set of deep adaptive-mesh-refinement simulations, we quantify the dependence of the SFR per free-fall time, $\epsilon_{\rm ff}$, on the above parameters. Our main results are: i) $\epsilon_{\rm ff}$ decreases exponentially with increasing $t_{\rm ff}/t_{\rm dyn}$, but is insensitive to changes in ${\cal M}_{\rm s}$, for constant values of $t_{\rm ff}/t_{\rm dyn}$ and ${\cal M}_{\rm a}$. ii) Decreasing values of ${\cal M}_{\rm a}$ (stronger magnetic fields) reduce $\epsilon_{\rm ff}$, but only to a point, beyond which $\epsilon_{\rm ff}$ increases with a further decrease of ${\cal M}_{\rm a}$. iii) For values of ${\cal M}_{\rm a}$ characteristic of star-forming regions, $\epsilon_{\rm ff}$ varies with ${\cal M}_{\rm a}$ by less than a factor of two. We propose a simple star-formation law, based on the empirical fit to the minimum $\epsilon_{\rm ff}$, and depending only on $t_{\rm ff}/t_{\rm dyn}$: $\epsilon_{\rm ff} \approx \epsilon_{\rm wind} \exp(-1.6 \,t_{\rm ff}/t_{\rm dyn})$. Because it only depends on the mean gas density and rms velocity, this law is straightforward to implement in simulations and analytical models of galaxy formation and evolution.


\end{abstract}


\section{Introduction}

Stars are the result of the gravitational collapse of cold gas, but on large scales the conversion of cold gas into stars takes
much longer than a collapse time. Early explanations for this low star formation rate (SFR) involved the magnetic support against the collapse
of giant molecular clouds (GMCs), while the more recent scenario of turbulent fragmentation relies on the support of GMCs by supersonic
turbulence \citep{maclow+klessen04}. However, numerical simulations of star formation in 
turbulent clouds have generally failed to yield SFRs as low as observed values\footnote{After submitting this Letter, we became aware of a 
closely related paper by Federrath and Klessen, submitted to the Astrophysical Journal (arXiv:1209.2856), presenting a large number of star formation 
simulations that are complementary to ours.} (see \S~4). 

This work shows that supersonic MHD turbulence results in a SFR as low as observed in GMCs, for characteristic values of 
the free-fall time divided by the dynamical time, $t_{\rm ff}/t_{\rm dyn}$, the alfv\'{e}nic Mach number, ${\cal M}_{\rm a}$, and the sonic Mach number, 
${\cal M}_{\rm s}$ (defined in \S~2). Using a very large set of adaptive-mesh-refinement (AMR) simulations of driven MHD turbulence, including self-gravity and sink 
particles, we quantify the dependence of the SFR per free-fall time, $\epsilon_{\rm ff}$, on $t_{\rm ff}/t_{\rm dyn}$, confirming the results of 
our previous uniform grid simulations in the case of a very weak mean magnetic field. 

The SFR per free-fall time is the efficiency factor of a theoretical Schmidt-Kennicut law, 
\begin{equation}
\dot{\rho}_{\rm stars}=\epsilon_{\rm ff} \rho_{\rm gas}/t_{\rm ff}, 
\label{sfr}
\end{equation}
often used in models and simulations of star formation in galaxies. We find that $\epsilon_{\rm ff}$ decreases exponentially with increasing 
$t_{\rm ff}/t_{\rm dyn}$, has no dependence on ${\cal M}_{\rm s}$ alone, and varies by less than a factor of two with variations in ${\cal M}_{\rm a}$ 
within reasonable values for star-forming regions. We therefore propose a new empirical law of star formation that depends only on $t_{\rm ff}/t_{\rm dyn}$.

\section{Numerical Simulations}

\begin{table}[t]\footnotesize
\caption{Non-dimensional parameters and $\epsilon_{\rm ff}$ of the simulations.}
\centering
\begin{tabular}{llccccccc}
\hline\hline \\[-2.2ex]
Run  & ${\cal M}_{\rm s}$ & ${\cal M}_{\rm a}$ & $t_{\rm ff}/t_{\rm dyn}$ & $L_{\rm J,i}/\Delta x_{\rm i}$  &  $\epsilon_{\rm ff}$
\\ [0.8ex]
\hline \\[-1.8ex]
M10B0.3G12.5    & 10 &  33  &  3.09  &   20                    &  0.00                     \\
M10B0.3G25       &  10 &  33  & 2.18  &   14                    &  0.10                      \\
M10B0.3G50       &  10 &  33  & 1.54  &   10                    &  0.20                      \\
M10B0.3G100     & 10 &   33  & 1.09  &   7                    &  0.41                      \\
M10B0.3G200     & 10 &   33  & 0.77  &   5                    &  0.46                      \\
M10B0.3G400     & 10 &   33  & 0.54  &   3.5                    &  0.59                      \\
\hline 
M10B2G12.5       & 10 &    5   & 3.09   &   20                    &  0.00                       \\
M10B2B25           & 10 &    5   & 2.18   &   14                    &  0.03                      \\
M10B2G50           & 10 &    5  & 1.54    &   10                    &  0.09                      \\
M10B2G100         & 10 &    5  & 1.09   &   7                     &  0.22                      \\
M10B2G200         & 10 &    5  & 0.77   &   5                     &  0.28                      \\
M10B2G400         & 10 &    5  & 0.54   &   3.5                     &  0.38                      \\
M10B2G800         & 10 &    5  & 0.39   &   2.5                     &  0.47                      \\
\hline\hline 
M20B1G50           &  20 &  20  & 3.09  &    10                    &  0.008                      \\
M20B1G100         &  20 &  20  & 2.18  &      7                    &  0.10                      \\
M20B1G200         & 20 &   20  & 1.54  &      5                   &  0.16                      \\
M20B1G400         & 20 &   20  & 1.09  &    3.5                    &  0.28                      \\
M20B1G800         & 20 &   20  & 0.77  &    2.5                    &  0.44                      \\
\hline 
M20B4G50           &  20 &  5   & 3.09  &    10                    &  0.006                      \\
M20B4G100         &  20 &  5   & 2.18  &      7                    &  0.04                      \\
M20B4G200         & 20 &   5   & 1.54  &      5                    &  0.10                      \\
M20B4G400         & 20 &   5   & 1.09  &   3.5                    &  0.20                      \\
M20B4G800         & 20 &   5   & 0.77  &   2.5                    &  0.28                      \\
\hline 
M20B16G50           &  20 &  1.25   & 3.09  &     10              &  0.008                      \\
M20B16G100         &  20 &  1.25   & 2.18  &       7              &  0.06                      \\
M20B16G200         & 20 &   1.25   & 1.54  &       5              &  0.15                      \\
M20B16G400         & 20 &   1.25   & 1.09  &    3.5              &  0.25                      \\
M20B16G800         & 20 &   1.25   & 0.77  &    2.5              &  0.30                       \\
\hline\hline
\multicolumn{6}{c}{Runs for study of numerical convergence:}        \\
\hline 
32M10B2G50          & 10 &    5  & 1.54    &   5                    &  0.07                    \\
128M10B2G50          & 10 &    5  & 1.54    &   20                    &  0.08                  \\
128M10B2G501e6           & 10 &    5  & 1.54    &   20                    &  0.07                      \\
32M10B2G100        & 10 &    5  & 1.09   &    3.5                     &  0.24                   \\
128M10B2G100         & 10 &    5  & 1.09   &    14                     &  0.22                     \\
128M10B2G1001e6  & 10 &    5  & 1.09   &    14                     &  0.22                       \\
32M10B2G800           & 10 &    5  & 0.39   &    1.25                     &  0.46                   \\
128M10B2G800         & 10 &    5  & 0.39   &      5                     &  0.49                \\
128M10B2G8001e6         & 10 &    5  & 0.39   &      5                     &  0.50        \\
\hline
\multicolumn{6}{c}{Runs for study of variations due to initial conditions:}        \\
\hline 
32M10B2G50b          & 10 &    5  & 1.54    &   5                    &  0.09                    \\
32M10B2G50c          & 10 &    5  & 1.54    &   5                    &  0.12                    \\
32M10B2G50d          & 10 &    5  & 1.54    &   5                    &  0.11                    \\
32M10B2G50e          & 10 &    5  & 1.54    &   5                    &  0.05                    \\
32M10B2G100b        & 10 &    5  & 1.09   &    3.5                     &  0.22                   \\
32M10B2G100c        & 10 &    5  & 1.09   &    3.5                     &  0.31                   \\
32M10B2G100d        & 10 &    5  & 1.09   &    3.5                     &  0.26                   \\
32M10B2G100e        & 10 &    5  & 1.09   &    3.5                     &  0.17                         
\\ [0.4ex]
\hline \\
\end{tabular}
\label{t1}
\end{table}

The simulations are carried out with the AMR code Ramses \citep{Teyssier02}, modified to include random turbulence 
driving and sink particles, and optimized for supercomputers with large multi-core nodes through an OpenMP/MPI hybrid layout. As 
in \citet{Padoan+Nordlund02imf,Padoan+Nordlund04bd,Padoan+Nordlund11sfr}, we adopt periodic boundary conditions, isothermal equation of state, 
and solenoidal random forcing in Fourier space at wavenumbers $1\le k\le2$ ($k=1$ corresponds to the computational box size). We choose a solenoidal 
force to make sure that dense, star-forming structures are naturally generated in the turbulent flow, rather than directly imposed by the external force. 

The simulations are first run without self-gravity for approximately five dynamical times, starting with uniform density and magnetic fields, and random
velocity field with power only at wavenumbers $1 \le k\le 2$. The driving force keeps the rms sonic Mach number,
${\cal M}_{\rm s}\equiv \sigma_{\rm v,3D}/c_{\rm s}$, at the approximate values of either 10 or 20, characteristic of GMCs on scales of order 10~pc 
($\sigma_{\rm v,3D}$ is the three-dimensional rms velocity, and
$c_{\rm s}$ is the isothermal speed of sound). For the simulations with  ${\cal M}_{\rm s}\approx 10$, we have chosen
two different values of the mean magnetic field, corresponding to rms alfv\'{e}nic Mach numbers of
${\cal M}_{\rm a}\approx 33$ and 5, where ${\cal M}_{\rm a}\equiv \sigma_{\rm v,3D}/v_{\rm a}$, and $v_{\rm a}$ is the Alfv\'{e}n velocity corresponding
to the mean magnetic field and the mean density. For the larger Mach number, ${\cal M}_{\rm s}\approx 20$, we have adopted three different values of the mean magnetic
field, corresponding to  ${\cal M}_{\rm a}\approx 20$, 5, and 1.25. 

In the supersonic regime, the turbulent dynamo is suppressed \citep{Haugen+04dynamo_mach}, and the magnetic energy growth 
saturates at approximately 2\% of the kinetic energy of the turbulence, when the mean magnetic field is negligible and does not 
characterize the system \citep{Federrath+11saturation}. On the other hand, if the mean magnetic field is not negligible, as in our simulations and in real GMCs, 
it is readily amplified, well beyond the dynamo saturation level, by stretching and compression in the turbulent flow, and thus its saturated rms value depends directly 
on its mean value, as found in simulations of \citet{Kritsuk+09}, and in equation (20) of \citet{Padoan+Nordlund11sfr}. Therefore, the value of ${\cal M}_{\rm a}$, 
defined with the mean magnetic field, characterizes the saturated regime and can be taken as a fundamental parameter of the system.

The final snapshots of these five turbulence simulations are used as initial conditions for the star-formation simulations with self-gravity and sink particles. 
Each initial condition is used for 5 to 7 simulations with different numerical values of the gravitational constant, corresponding to different values of $t_{\rm ff}/t_{\rm dyn}$, 
resulting in 28 star-formation simulations (plus 6 larger runs used to test numerical convergence, and 11 smaller runs to test both convergence and variations of 
$\epsilon_{\rm ff}$ with initial conditions -- a total of 45 star-formation simulations). The dynamical time is defined as $t_{\rm dyn}\equiv L/ 2 \sigma_{\rm v,3D}$, where
$L$ is the size of the computational domain, and $t_{\rm ff}\equiv \sqrt{3 \pi/32 G \rho_0}$, where $\rho_0$ is the mean density in the computational volume.
The simulations were run until the star formation efficiency value of SFE$\,\, \equiv M_{\star}/M\,\approx$ 0.2-0.3, where $M_{\star}$ is the mass in the sink particles, and $M$ the total
mass (gas plus sinks), except the simulations with very low SFR, that were run until SFE$\,\approx$ 0.05-0.1. 

The five initial turbulence simulations have a root grid of 128$^3$ computational cells and 5 AMR levels, thus a maximum resolution equivalent to 4,096$^3$ cells.
In all runs, the first AMR level is created if the density is 2.5 times larger than the mean. Subsequent AMR levels are created when the 
gas density increases by a factor of 4 relative to the previous level. In contrast to \citet{Kritsuk+06} and \citet{Schmidt+09}, we do not refine based on velocity or pressure 
gradients. However, we verified that the tails of the density PDFs are sufficiently converged in the turbulence simulations used as initial conditions (details will be given 
elsewhere). When self-gravity is included, the root grid is reduced to 64$^3$ cells, while the number of AMR levels is increased from 5 to 8, to achieve a maximum 
resolution equivalent to 16,384$^3$ cells. 

Sink particles are created in cells where the gas density is larger than $10^5$ times the mean density (much larger than the maximum density reached in the initialization 
runs without self-gravity), if the following additional conditions are met at the cell location: i) The gravitational potential has a local minimum value, ii) the three-dimensional
velocity divergence is negative, and iii) no other previously created sink particle is present within an exclusion radius, $r_{\rm excl}$ ($r_{\rm excl}=17.5 \Delta x$ in these 
simulations, where $\Delta x$ is the highest-resolution cell size). These conditions are similar to those in \citet{Federrath+10sinks}. We have verified that they avoid 
the creation of spurious sink particles in regions where the gas is not collapsing.

\begin{figure}[t]
\includegraphics[width=\columnwidth]{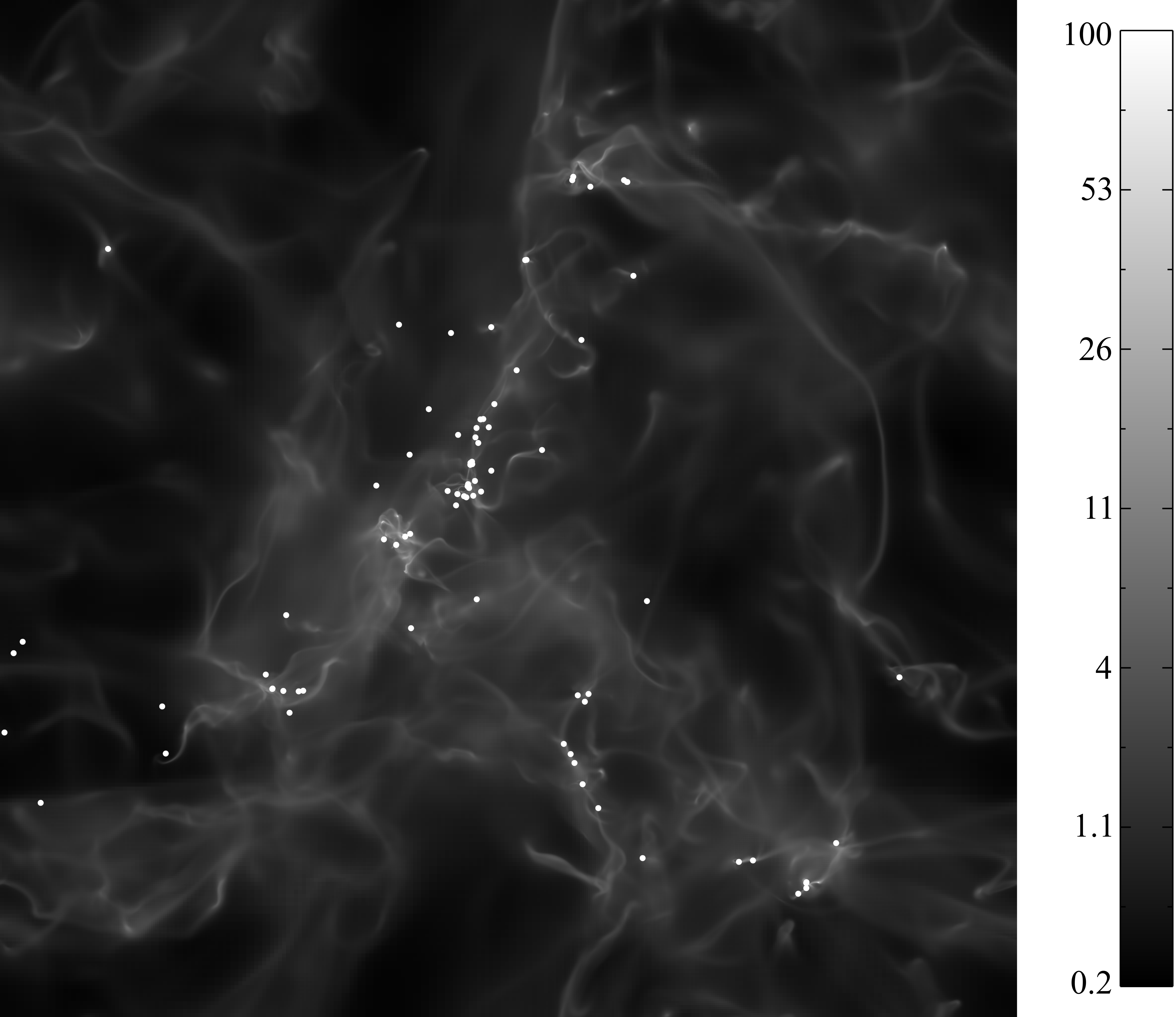}
\caption[]{Projected density, $N_{\rm col}$, and sink particles from the run 128M10B2G50, at a time when SFE$\approx0.05$. The color scale is linear in $N_{\rm}^{0.2}$,
and the tick values in the color bar give $N_{\rm col}$ in numerical units, with mean gas density and computational box size equal to unity.}
\label{}
\end{figure}

The numerical parameters characterizing the simulations are given in Table~1. The simulation names contain the rms sonic Mach number, and the numerical values 
of the mean magnetic field and of the gravitational constant. In Table~1, we give three non-dimensional parameters,  
${\cal M}_{\rm s}$, ${\cal M}_{\rm a}$, and $t_{\rm ff}/t_{\rm dyn}$, that fully characterize the simulations. 
The ratio $t_{\rm ff}/t_{\rm dyn}$ expresses the relative strength of gravity and turbulence, like the virial parameter, $\alpha_{\rm vir}$, used in previous works. 
Although formally equivalent to $\alpha_{\rm vir}$, for example through the usual expressions valid for a uniform sphere, we prefer to use $t_{\rm ff}/t_{\rm dyn}$ 
as the third parameter, in line with the choice of the Mach numbers (ratios of crossing times as well) as the first two parameters. 

In Table~1, we also give the ratio 
between the minimum Jeans length at each level, $L_{\rm J,i}$ (corresponding to the highest density before further refinement), and the computational 
cell size, $\Delta x_{\rm i}$, at that level. Due to the refinement by a factor of two in size for every increase by a factor of four in density, the minimum number of cells 
per Jeans length, $L_{\rm J,i}/\Delta x_{\rm i}$, is the same for every level $i$ in a given simulation. The last column of Table~1 shows the values of $\epsilon_{\rm ff}$ 
obtained from the simulations. 

At the bottom of Table~1, we also list the simulations used to test numerical convergence and to study the variations of 
$\epsilon_{\rm ff}$ with initial conditions. The names of these extra runs begin with their root grid size, either $128^3$ or $32^3$ computational cells.
Projected density and sink particles from one of these runs, 128M10B2G50, are shown in Figure~1, at a time when SFE$\approx0.05$.

\begin{figure}[t]
\includegraphics[width=\columnwidth]{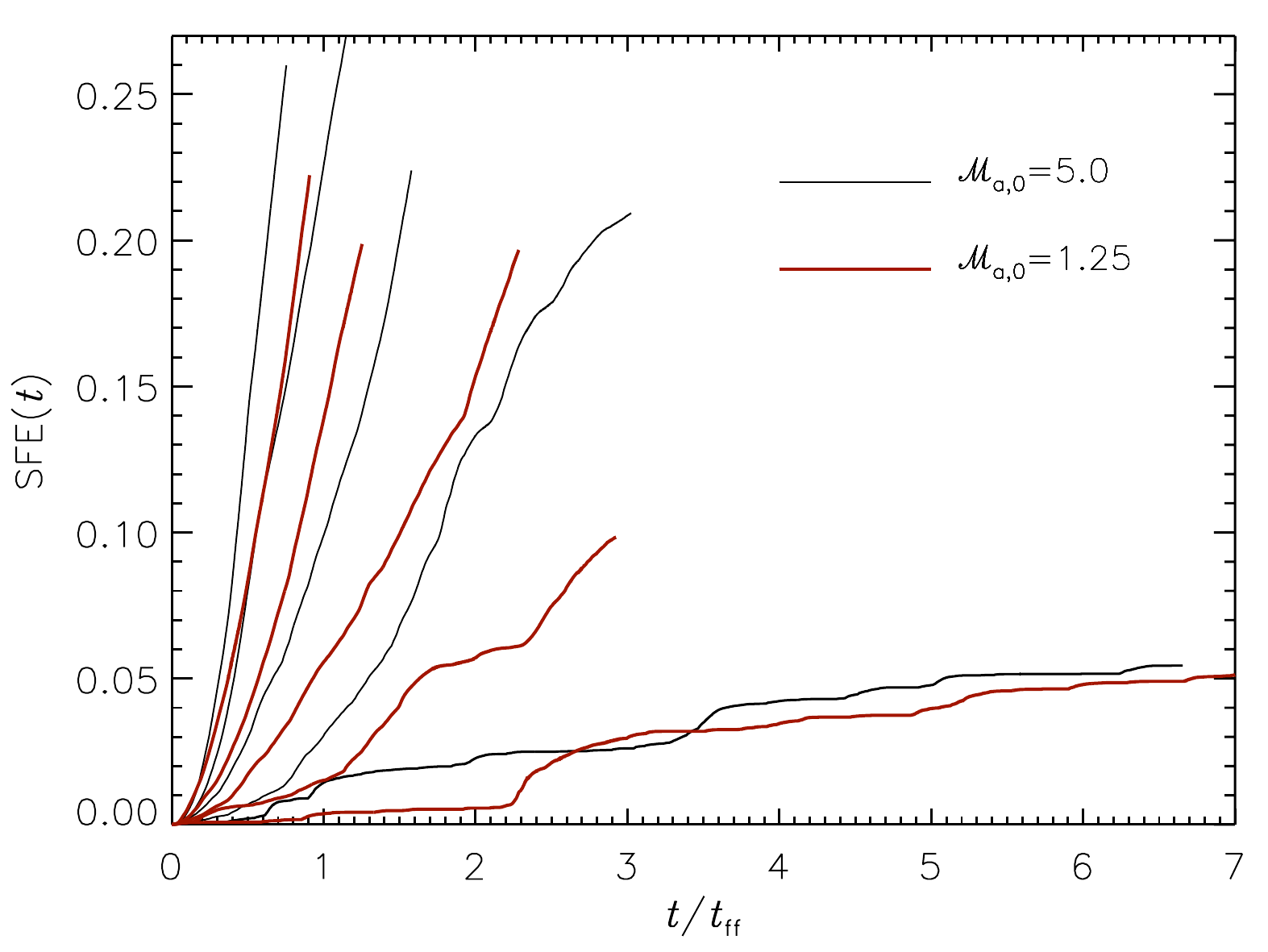}
\caption[]{SFE versus time for the two sets of simulations with ${\cal M}_{\rm s}\approx 20$ and ${\cal M}_{\rm a}\approx 5$ and 1.25. The values of 
$t_{\rm ff}/t_{\rm dyn}$ are 0.77, 1.09, 1.54, 2.18, and 3.09, from top (high SFE) to bottom.}
\label{}
\end{figure}

\section{Results}

Figure~2 shows the time dependence of the SFE for the two sets of simulations with ${\cal M}_{\rm s}\approx 20$ and ${\cal M}_{\rm a}\approx 5$ and 1.25.
In some runs with high $\epsilon_{\rm ff}$, the SFR tends to grow with time. To extract a representative value, we measure $\epsilon_{\rm ff}$ as the slope of 
the least-squares fit of the SFE versus $t/t_{\rm ff}$, in the interval $0.03 \le {\rm SFE} \le 0.2$, as in \citet{Padoan+Nordlund11sfr}. 
The simulations with the lowest values of $\epsilon_{\rm ff}$ were not integrated up to SFE=0.2. However, they do not exhibit a systematic 
growth of the SFR with time, and they are run for many free-fall times (up to 9 in some cases), so their average $\epsilon_{\rm ff}$ is nonetheless well defined.

The values of $\epsilon_{\rm ff}$ from all simulations are plotted in Figure~3, as a function of $t_{\rm ff}/t_{\rm dyn}$. A striking result, stressed by the linear-log 
scale of Figure~3, is the approximately exponential dependence of $\epsilon_{\rm ff}$ on $t_{\rm ff}/t_{\rm dyn}$ (for fixed values of ${\cal M}_{\rm s}$ and 
${\cal M}_{\rm a}$), showing that supersonic MHD turbulence is very effective at slowing down the star formation process, which seems to be controlled 
primarily by the ratio of the characteristic times of gravity and turbulence.

Another important result is the lack of a dependence of $\epsilon_{\rm ff}$ on ${\cal M}_{\rm s}$, for fixed values of ${\cal M}_{\rm a}$ and $t_{\rm ff}/t_{\rm dyn}$. 
Furthermore, the dependence on ${\cal M}_{\rm a}$ is not very strong, considering the very large range of ${\cal M}_{\rm a}$ values of the simulations. 
This is in part due to the complex dependence of $\epsilon_{\rm ff}$ on ${\cal M}_{\rm a}$. 
Down to ${\cal M}_{\rm a}\approx 5$, $\epsilon_{\rm ff}$ decreases with decreasing ${\cal M}_{\rm a}$ (increasing magnetic field strength), while for further 
decreasing ${\cal M}_{\rm a}$, $\epsilon_{\rm ff}$ starts to increase. 

Figure~3 shows that a simple exponential law, where $\epsilon_{\rm ff}$ is only controlled by the ratio $t_{\rm ff}/t_{\rm dyn}$, could describe all 
simulation results within a factor of two. For even weaker magnetic fields than considered here, $\epsilon_{\rm ff}$ would certainly be larger, as it 
should approach the values found in simulations without magnetic fields \citep{Padoan+Nordlund11sfr}. However, observations in star forming regions
are consistent with ${\cal M}_{\rm a}\approx 1-10$ \citep{Padoan+Nordlund99MHD, Lunttila+08, Lunttila+09, Heyer+Brunt12taurus}, so it is likely that 
$\epsilon_{\rm ff}$ approaches its minimum value, which is approximately fit by a simple exponential law, $\epsilon_{\rm ff}\approx \exp(-1.6 \, t_{\rm ff}/t_{\rm dyn})$, 
shown as a dashed line in Figure~3. We therefore propose a new empirical law of star formation, based on this fit to the minimum $\epsilon_{\rm ff}$:
\begin{equation}
\epsilon_{\rm ff} \approx \epsilon_{\rm wind} \exp(-1.6 \,t_{\rm ff}/t_{\rm dyn}),
\label{sfrff}
\end{equation}
where $\epsilon_{\rm wind}\approx 0.5$ accounts for mass loss through jets and winds, during the formation of a star. This law only depends on the mean 
gas density and the rms velocity of a star-forming region, so it is easily implemented in analytical models and simulations of galaxy formation and evolution.

\begin{figure}[t]
\includegraphics[width=\columnwidth]{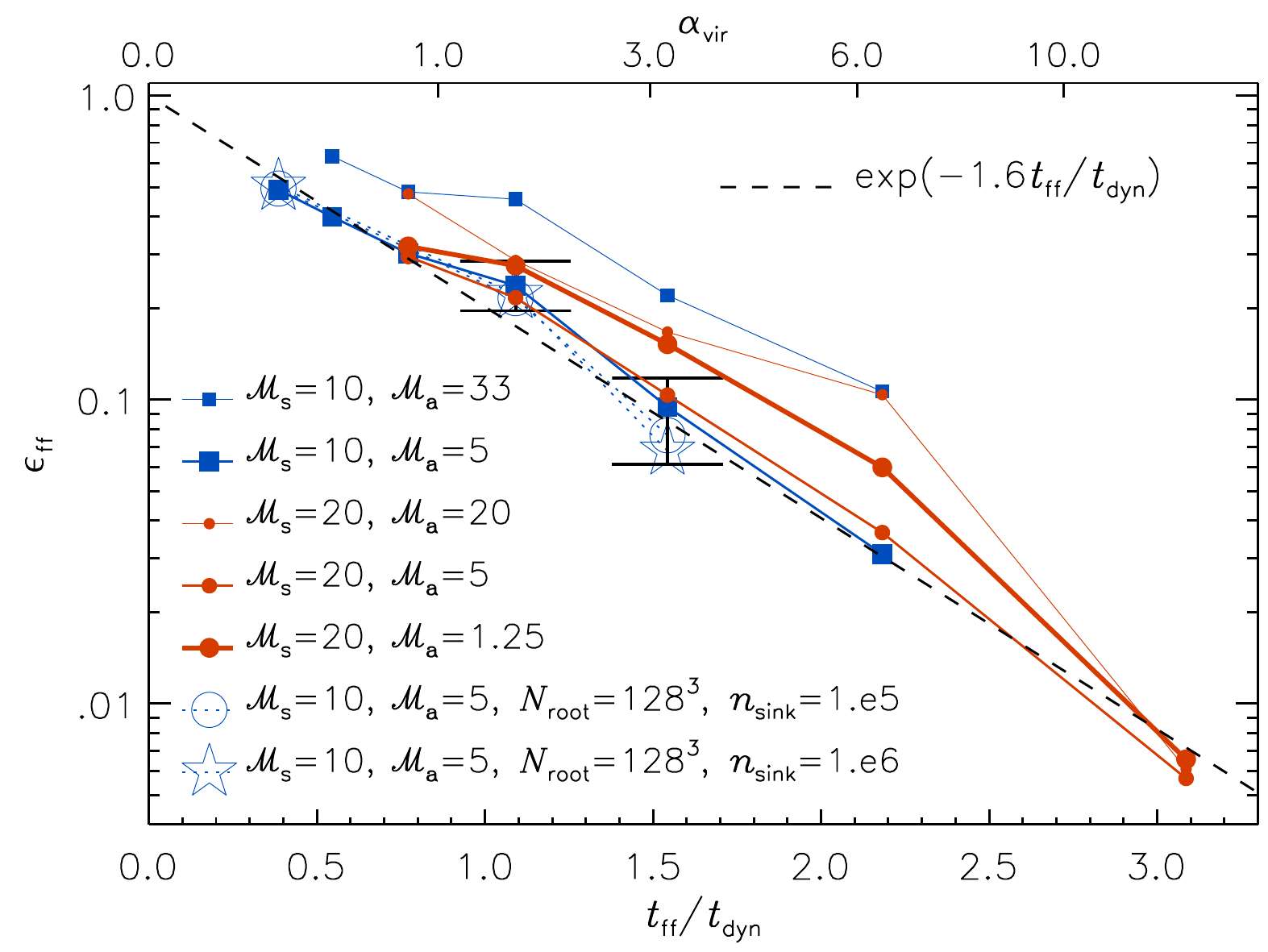}
\caption[]{SFR per free-fall time, $\epsilon_{\rm ff}$, versus $t_{\rm ff}/t_{\rm dyn}$. The symbols for each series of runs, where
only the strength of gravity is changed and ${\cal M}_{\rm s}$ and ${\cal M}_{\rm a}$ are kept constant, are connected by a line, to better distinguish 
each series. The two error bars give the mean of $\epsilon_{\rm ff}$, plus and minus the rms values, for each group of five $32^3$-root-grid runs 
with identical parameters (${\cal M}_{\rm s}\approx 10$ and ${\cal M}_{\rm a}\approx 5$), but different initial conditions. The dashed line is an approximate 
exponential fit to the minimum value of $\epsilon_{\rm ff}$ versus $t_{\rm ff}/t_{\rm dyn}$.}
\label{}
\end{figure}

Equation (\ref{sfrff}) is valid as long as the fraction of the system mass, $M_0$, 
available to form stars is larger than the characteristic stellar
mass, $m_{\star}$, which we can express as $\epsilon_{\rm ff} M_0 /m_{\star} >1$. If $\epsilon_{\rm ff}$ predicted by equation (\ref{sfrff}) is
too small to satisfy this condition, then $\epsilon_{\rm ff}=0$, simply because the total system mass is too small. Adopting the characteristic stellar mass predicted 
by our turbulent fragmentation model \citep{Padoan+Nordlund02imf, Padoan+07imf}, $m_{\star}\approx 3 M_{\rm BE,0}/ {\cal M}_{\rm s}$, 
which corresponds to approximately three times the mean Bonnor-Ebert mass, $M_{\rm BE,0}$, evaluated for an external pressure equal 
to the ram pressure of the turbulence, and using the minimum $\epsilon_{\rm ff}$ derived here, we find the condition can be expressed as
$\epsilon_{\rm ff} M_0 /m_{\star} \approx \epsilon_{\rm ff} (t_{\rm ff}/t_{\rm dyn})^{-3}{\cal M}_{\rm s}^4>1$. 

As shown by Table~1, the weakest gravity we have simulated corresponds to $t_{\rm ff}/t_{\rm dyn}\approx 3.09$, giving 
$\epsilon_{\rm ff} M_0/m_{\star}\approx 2.4$ and 38.6 for the runs with ${\cal M}_{\rm s}\approx 10$ and 20 respectively.
In good agreement with these values, in each of the ${\cal M}_{\rm s}\approx 10$ runs with $t_{\rm ff}/t_{\rm dyn}\approx 3.09$, 
we get only a single star, so the SFR is practically zero, while in the three ${\cal M}_{\rm s}\approx 20$ runs with $t_{\rm ff}/t_{\rm dyn}\approx 3.09$ 
we get approximately 10-20 stars per free-fall time. 

To test the numerical convergence of $\epsilon_{\rm ff}$, we have rerun three of the simulations
with ${\cal M}_{\rm s}\approx 10$ and ${\cal M}_{\rm a}\approx 5$, using both twice smaller and twice larger root grid size ($32^3$ and $128^3$
computational cells), and the same number of AMR levels, corresponding to maximum resolutions 
equivalent to 8,192$^3$ cells and 32,768$^3$ cells. These three larger runs were then rerun again using a 10 times larger
value of the threshold density above which sink particles are created (10$^6$ times the mean density, instead of $10^5$). 
As shown in Table~1, the values of  $\epsilon_{\rm ff}$ in the runs with $t_{\rm ff}/t_{\rm dyn}=0.39$ and 1.09 are practically independent of numerical resolution,
showing that the correct SFR can be derived even when Truelove's condition \citep{truelove.....97} is violated, and that the value of
$10^5$ times the mean density (adopted in the main parameter study), above which sink particles are allowed to form, is sufficiently high to avoid dense 
regions that are not collapsing.  

In the case of $t_{\rm ff}/t_{\rm dyn}=1.54$, $\epsilon_{\rm ff}$ is approximately 20\% lower in the $128^3$-root-grid simulations than in the corresponding
$64^3$-root-grid run. However, this low SFR is obtained from an average over many dynamical times. During the first dynamical time, the 
SFR is virtually independent of resolution. The corresponding simulation with $32^3$ root-grid size has the same value of $\epsilon_{\rm ff}$
as the largest simulations, showing that the difference from the $64^3$-root-grid run is not due to lack of convergence, but more likely to 
random differences between runs after a long time, as expected in chaotic systems. The values of $\epsilon_{\rm ff}$ from the $128^3$-root-grid 
simulations can be seen in Figure~3.

To investigate the effect of the initial conditions, we have carried out 10 simulations with twice smaller root grid size ($32^3$ computational cells), 
and the same number of AMR levels, five of them with $t_{\rm ff}/t_{\rm dyn}=1.09$, and the other five with $t_{\rm ff}/t_{\rm dyn}=1.54$, starting from 
different initial conditions of fully-developed turbulence, with ${\cal M}_{\rm s}\approx 10$ and ${\cal M}_{\rm a}\approx 5$. As shown in Table~1, 
the maximum difference in $\epsilon_{\rm ff}$ from such small samples is approximately a factor of two. The mean values of $\epsilon_{\rm ff}$, 
plus and minus the rms values, are plotted as error bars in Figure~3.

\section{Discussion}

The values of $\epsilon_{\rm ff}$ predicted by equation (\ref{sfrff}) are consistent with those found in GMCs. As discussed in \citet{Padoan+Nordlund11sfr},
accounting for the full uncertainty in the lifetime of the Class II phase ($2\pm1$~Myr), the results of \citet{Evans+09} implies $\epsilon_{\rm ff}=0.02$-0.12. 
Using the GMC sample by \citet{Heyer+09}, we derive $t_{\rm ff}/t_{\rm dyn}\approx 1.4 \pm 0.5$
(1-$\sigma$), resulting in $\epsilon_{\rm ff}=0.02$-0.12 based on our equation  (\ref{sfrff}), with $\epsilon_{\rm wind}\approx 0.5$, in nice agreement with 
the values derived by \citet{Evans+09} (or $t_{\rm ff}/t_{\rm dyn}\approx 1.0 \pm 0.4$ and $\epsilon_{\rm ff}=0.06$-0.19 if we double the cloud masses). 
Similar values are found also for the Orion Nebula Cluster, $\epsilon_{\rm ff}=0.03$-0.09, the object with the smallest error bars in \cite{Krumholz+Tan07slowsf}.  
Based on the values of $t_{\rm ff}/t_{\rm dyn}$ in the sample of \citet{Murray11} we would predict an average value of $\epsilon_{\rm ff}=0.13\pm 0.07$, 
while, based on the free-free emission, Murray finds an average value of $\epsilon_{\rm ff}=0.14\pm 0.14$.

The AMR simulations with ${\cal M}_{\rm s}\approx 10$ and ${\cal M}_{\rm a}\approx 33$ presented here are consistent with our previous uniform grid 
simulations \citep{Padoan+Nordlund11sfr} with ${\cal M}_{\rm s}\approx 9$ and ${\cal M}_{\rm a}\approx 30$ (the values of $\alpha_{\rm vir}$ given in the 
plot of Figure~2 are derived from $t_{\rm ff}/t_{\rm dyn}$ based on the case of a uniform sphere, while those reported in Figure~7 of \citet{Padoan+Nordlund11sfr} 
were based on the total mass in the computational box, and should be multiplied by a factor of 2 for the comparison).

The SFR model of \citet{Padoan+Nordlund11sfr} adopted the simplifying assumption that, in the MHD case, the effective postshock ratio of gas to magnetic pressure, 
$\beta$, and the density pdf do not depend on the mean magnetic field strength, for reasonable values of ${\cal M}_{\rm a}$. The complex, but weak dependence of 
$\epsilon_{\rm ff}$ on ${\cal M}_{\rm a}$ found in the simulations justifies that approximation. However, by neglecting the dependence of $\beta$ on both 
${\cal M}_{\rm a}$ and ${\cal M}_{\rm s}$, the model yields a direct dependence on ${\cal M}_{\rm s}$, not found in our AMR simulations. Equation (20) of 
\citet{Padoan+Nordlund11sfr}, relating $\beta$ to ${\cal M}_{\rm s}$ and $\beta_0$, the ratio of gas to magnetic pressure computed with the average magnetic 
field strength, $\beta \propto \beta_0^{1/2}{\cal M}_{\rm s}^{-1}$,  was derived under the assumption of flux freezing, neglecting the postshock thermal pressure 
and dynamical alignment of magnetic and velocity fields. \citet{Padoan+Nordlund11sfr} noticed that when $\beta$ is averaged at high densities, it becomes 
independent of $\beta_0$, so they took $\beta=const$, which also provided a good fit to the density pdfs. However, they did not test the dependence of $\beta$ on ${\cal M}_{\rm s}$. 

It may be more physical to retain the dependence of $\beta$ on ${\cal M}_{\rm s}$ even at high densities, because the postshock magnetic pressure 
tends to balance the ram pressure of the turbulence causing the shocks. Assuming $\beta \propto {\cal M}_{\rm s}^{-1}$ 
instead of $\beta=0.39$, we obtain $\epsilon_{\rm ff}$ practically independent of ${\cal M}_{\rm s}$, as found in our AMR simulations. This assumption yields a density 
variance in MHD given by $\sigma_{\rm x,MHD}\approx {\cal M}_{\rm s}^{1/2}/2=\sigma_{\rm x,HD}/{\cal M}_{\rm s}^{1/2}$, for ${\cal M}_{\rm s}\gg 1$, while the 
approximation $\beta=0.39$ in \citet{Padoan+Nordlund11sfr} gave $\sigma_{\rm x,MHD}\approx 0.53{\cal M}_{\rm s}/2 = 0.53 \sigma_{\rm x,HD}$ (this issue will be
addressed in a future study).

Although this correction eliminates the dependence on ${\cal M}_{\rm s}$, the model does not predict the subtle dependence of $\epsilon_{\rm ff}$ on ${\cal M}_{\rm a}$, 
as it does not account for the increasing anisotropy and dynamical alignment of magnetic and velocity fields, with decreasing ${\cal M}_{\rm a}$. Therefore,
the simple empirical law derived above is a better choice for practical applications. 

The models by \citet{Krumholz+McKee05sfr} and \cite{Hennebelle+Chabrier11sfr} do not include the effect of magnetic fields. As shown by our previous 
uniform-grid simulations \citep{Padoan+Nordlund11sfr}, even with a very weak mean magnetic field, corresponding to ${\cal M}_{\rm a}\approx 30$, 
$\epsilon_{\rm ff}$ is approximately three times smaller than in the non-magnetized case. The AMR simulations presented here show that, 
with more realistic, larger magnetic field values, $\epsilon_{\rm ff}$ decreases even further.
Analytical models of star formation neglecting magnetic fields should not yield low values of $\epsilon_{\rm ff}$ consistent with observations 
of star-forming regions. If they do, they are inconsistent with the numerical experiments.

\section{Conclusions}

The new AMR simulations of star formation presented in this work show that i) $\epsilon_{\rm ff}$ decreases exponentially with increasing $t_{\rm ff}/t_{\rm dyn}$, 
but is insensitive to changes in ${\cal M}_{\rm s}$ (in the range $10\le{\cal M}_{\rm s}\le20$), for constant values of $t_{\rm ff}/t_{\rm dyn}$  and ${\cal M}_{\rm a}$. ii) Decreasing values of 
${\cal M}_{\rm a}$ (increasing magnetic field strength) reduce $\epsilon_{\rm ff}$, but only to a point, beyond which $\epsilon_{\rm ff}$ increases with a further 
decrease of ${\cal M}_{\rm a}$. iii) For values of ${\cal M}_{\rm a}$ characteristic of star-forming regions, $\epsilon_{\rm ff}$ varies with ${\cal M}_{\rm a}$ by less 
than a factor of two. We propose a simple law of star formation depending only on $t_{\rm ff}/t_{\rm dyn}$, based on the empirical fit to the minimum $\epsilon_{\rm ff}$: 
$\epsilon_{\rm ff} \approx \epsilon_{\rm wind} \exp(-1.6 \,t_{\rm ff}/t_{\rm dyn})$. 

This law shows that MHD turbulence is very effective at slowing down the star-formation process and can explain the 
low average SFR in molecular clouds. Because it only depends on the mean gas density and rms velocity of a star-forming region, the star-formation 
law we propose is straightforward to implement in simulations and analytical models of galaxy formation and evolution. Future work should test its effect 
in simulations where the feedback of SNs is accounted for, but the process of star formation is not spatially resolved and needs to be modeled.

\acknowledgements

We thank Christoph Federrath and the referee for useful comments on the manuscript.
PP is supported by the Spanish MICINN grant AYA2010-16833 and by the FP7-PEOPLE-2010-RG grant PIRG07-GA-2010-261359. 
TH and {\AA}N are supported by the Danish National Research Foundation, through its establishment of the Centre for Star and Planet Formation.
The simulations were carried out on the NASA/Ames Pleiades supercomputer.

\end{document}